\documentclass[aps,prl,eprint,groupaddress]{revtex4-1}
\def\Vec#1{\mbox{\boldmath $#1$}}
\usepackage[dvips]{graphicx}
\begin{document}

\title{Circularly-Polarized Light Emission from Semiconductor Planar Chiral Photonic Crystal}

\author{Kuniaki~Konishi$^{1,2,3}$, Masahiro~Nomura$^4$, Naoto~Kumagai$^4$, Satoshi~Iwamoto$^{4,5}$, Yasuhiko~Arakawa$^{4,5}$, and Makoto Kuwata-Gonokami$^{1,2,3,5}$}
\email{gonokami@ap.t.u-tokyo.ac.jp}
\affiliation{$^1$Photon Science Center, The University of Tokyo, Tokyo 113-8656, Japan\\
$^2$Department of Applied Physics, The University of Tokyo, Tokyo 113-8656, Japan\\
$^3$Core Research for Evolutional Science and Technology (CREST), JST, Tokyo 113-8656, Japan\\
$^4$Institute for Nano Quantum Information Electronics, The University of Tokyo, Tokyo 153-8505, Japan\\
$^5$Institute of Industrial Science, The University of Tokyo, Tokyo 153-8505, Japan}

\date{\today}

\begin{abstract}
We proposed and demonstrated a scheme of surface emitting circularly polarized light source by introducing strong imbalance between left- and right-circularly polarized vacuum fields in an on-waveguide chiral grating structure. We observed circularly polarized spontaneous emission from InAs quantum dots embedded in the wave guide region of a GaAs-based structure. Obtained degree of polarization reaches as large as 25\% at room temperature. Numerical calculation visualizes spatial profiles of the modification of vacuum field modes inside the structure with strong circular anisotropy.
\end{abstract}
\maketitle

The polarization and intensity of the emitted light depends on internal structure of the light source as well as on the symmetry and density of environment-allowed electromagnetic modes. The role of the environment in the light emission can be visualized by placing a light emitter in a microcavity \cite{16,17,18,19,20} because modification - in comparison with free space - in the allowed electromagnetic field modes affects the spontaneous emission rate. The modification of the mode structure can also affect radiation pattern and direction \cite{21,22} of the emitted light. The polarization sensitivity of the spontaneous emission enables control the polarization plane azimuth of a surface emitting device \cite{23,24}, and also has an important fundamental aspect. Specifically, broken time-reversal symmetry results in different emission efficiency for left- and right-circularly polarized photons in presence of a static magnetic field \cite{11,12,13,14,15}. One may expect that similar imbalance between the left- and right-circularly polarized photons should occur when left- and right-circularly polarized electromagnetic modes of the vacuum field are not equivalent. 

Circularly-polarized light is also  great important for a variety of applications, such as circular dichroism spectroscopy \cite{1} and chiral synthesis \cite{2,3} in biology and chemistry, spin-state control in quantum information technology \cite{4,5} and ultrafast magnetization control \cite{6,7}. However, despite of recent advances in the nanoscale lasing \cite{8,9,10}, circularly-polarized light sources fabricated with conventional semiconductor material that can be incorporated into optoelectronic circuits has not been realized yet.

Here, we present the circularly-polarized light emitter based on the control of the balance between left-and right-circularly polarized vacuum electromagnetic modes in  semiconductor chiral photonic crystal. We design and fabricate a GaAs-based semiconductor chiral photonic crystal with incorporated InAs quantum dot (QD) layer that produces circularly-polarized photoluminescence at room temperature. The performed numerical simulation allowed us to visualize the asymmetry of the vacuum modes coupled with circularly-polarized light propagating along the surface normal.

 \begin{figure}
 \includegraphics[width=85mm]{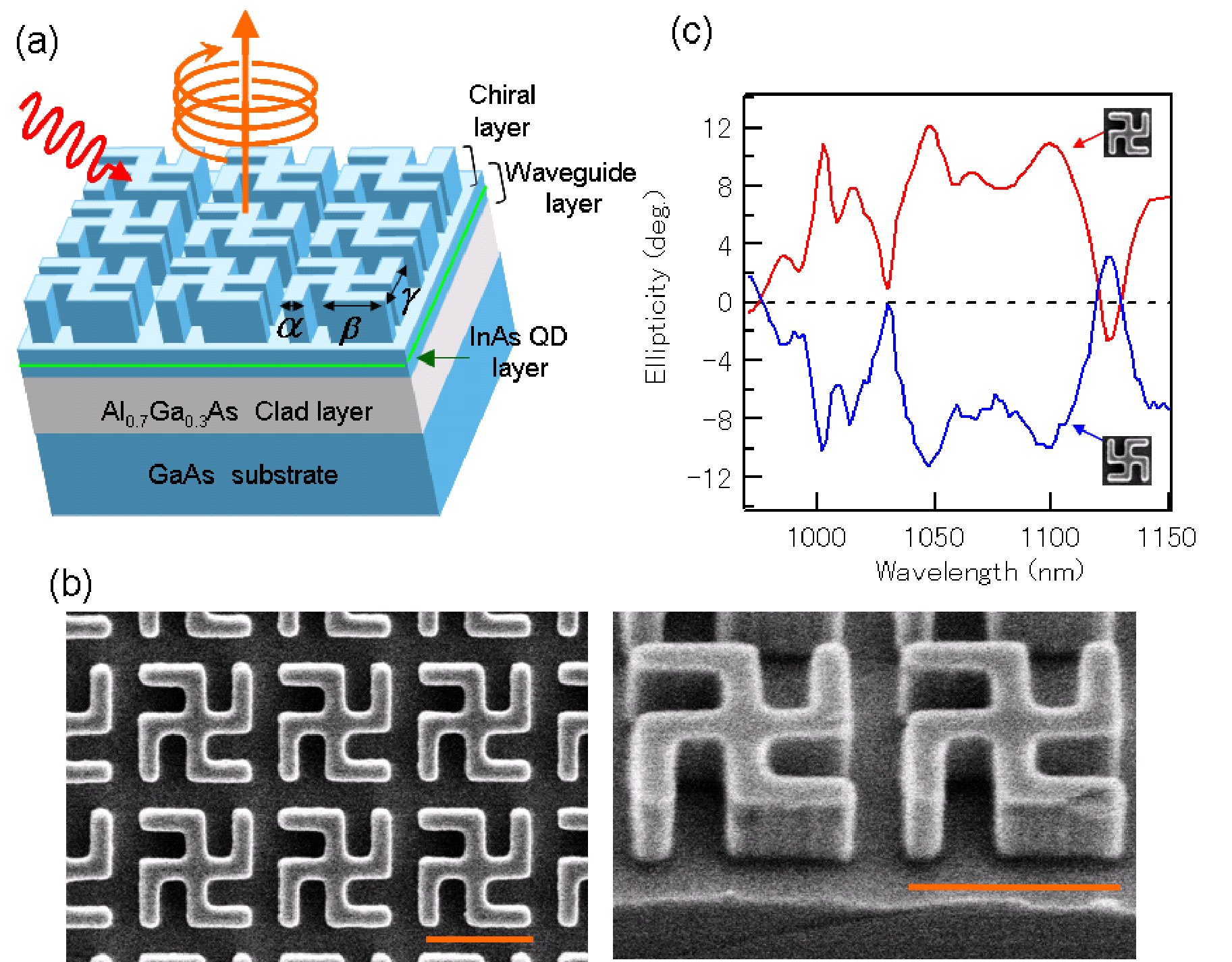}
 \caption{(a) Schematic figure of designed chiral photonic crystals. Chiral layer and waveguide layer are fabricated with GaAs. The thickness of the chiral layer, waveguide layer and clad layer are 460nm, 240nm and 1$\mu$m, respectively. The InAs QD single layer is placed in the middle of the waveguide layer. The value of the lengths $\alpha$, $\beta$, $\gamma$ are 153nm, 579nm, 1067nm, respectively. The period of the chiral grating is 1278nm~1278nm. (b) SEM images of a chiral photonic crystal with left-twisted gammadion (top- and bird's eye- view). The length of the orange scale-bars in the figures corresponds to 1$\mu$m. (c) Measured ellipticity spectra in the zero-order transmitted light of the chiral photonic crystals. Red (blue) line indicates the result obtained with the left- (right-)twisted gammadion sample. }
 \end{figure}

As shown in Fig. 1(a), the chiral photonic crystal consists of GaAs chiral nanograting layer, GaAs waveguide layer incorporating an emitter layer with InAs QDs and Al$_{0.7}$Ga$_{0.3}$As clad layer on GaAs substrate. The GaAs chiral nanograting layer is composed of gammadions that have no in-plane mirror symmetry but possess four-fold rotational axis. Such planar chiral structures possess strong optical activity \cite{25,26,27} and their handedness is determined by the sense of twist of gammadions. The Al$_{0.7}$Ga$_{0.3}$As layer (1$\mu$m), lower GaAs layer (150nm) including single InAs self-assembled QD layer, upper GaAs layer (550nm) are grown on GaAs(100) substrate by molecular beam epitaxy. The density of the QDs is 7~10$^9$/cm$^2$. The gammadion structures were patterned using an electron-beam lithography system and then transferred to the GaAs layer by inductive coupled plasma reactive ion etching (ICP-RIE) using a Cl$_2$/Ar mixture \cite{30}. The gammadion array is arranged along $\left\langle 110 \right\rangle$ and $\langle$\mbox{1-10}$\rangle$ direction of GaAs crystal. The etching times are controlled to achieve the desired thickness of the chiral layer. The whole size of the sample is 1.2mm~1.2mm. Figure 1(b) shows Scanning Electron Microscope (SEM) images of a fabricated left-twisted structure.

We elucidate the chirality of the manufactured structures by measuring the polarization plane azimuth rotation $\theta$ and ellipticity $\eta$ of a transmitted light wave at normal incidence in a wavelength range from 900nm to 1600nm with conventional polarization modulation technique \cite{31} (details of measurement are shown in Ref. \cite{32}). Although transmission spectra for both left- and right-twisted gammadion structures are nearly identical, signs of both $\theta$ and $\eta$ are opposite at all wavelengths. Fig. 1(c) shows ellipticity spectrum in InAs QDs luminescence spectrum range for left- and right-twisted gammadions.

\begin{figure}
 \includegraphics[width=85mm]{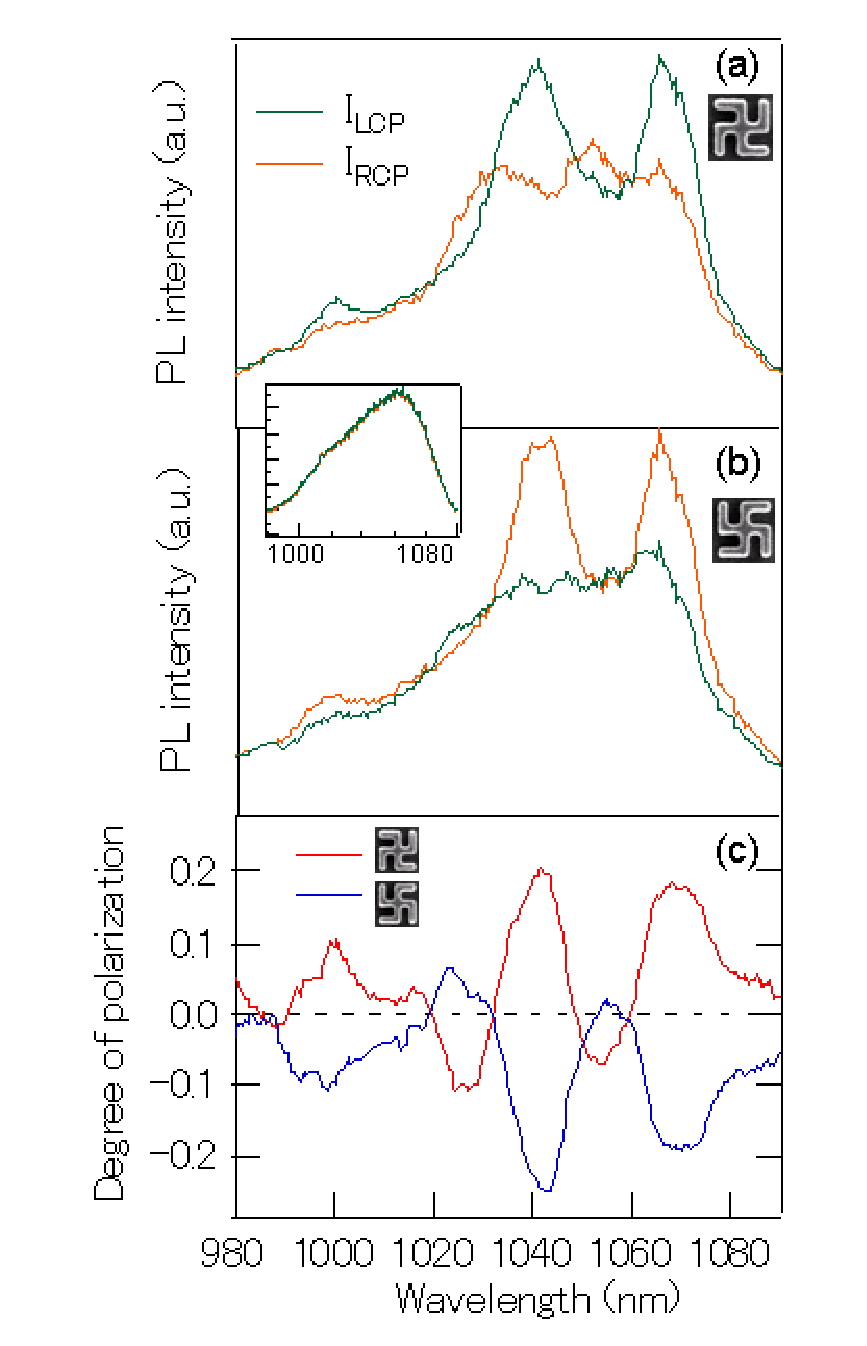}
 \caption{PL spectra of left(green line, $I_{LCP}$) and right(orange line, $I_{RCP}$) -circularly polarized components from the left-twisted (a) and right-twisted (b) chiral photonic crystals. Inset figure shows The PL spectra from InAs QDs without nanostructures. (c)Spectra of the degree of polarization of the chiral photonic crystals with the left-(red line) and right-(blue line) gammadions. }
 \end{figure}

	We measure the photoluminescence (PL) of manufactured chiral photonic crystals along the sample normal (N.A.`0.03). We performed photoluminescence measurement using a 635nm CW laser diode as an excitation source. The power and diameter of the beam spot at the surface of the samples are 2.6mW and 0.6mm, respectively. We distinguished between left- and right-circularly polarized components by setting a $\lambda /4$ waveplate and a polarizer in front of a detector and rotating the $\lambda /4$ waveplate. All measurements are performed at room temperature.

One can observe from Fig. 2(a) that PL spectra of the photonic crystals with left-twisted gammadion differ considerably for left- ($I_{LCP}$) and right-circularly ($I_{RCP}$) polarized components. By comparison, the PL spectra of InAs QDs without chiral nanostructures are shown in the inset of Fig. 2(a). In particular, a large difference between $I_{LCP}$ and $I_{RCP}$ is observed around 1040nm, i.e. at wavelength where large ellipticity is observed in the transmission measurement (see Fig. 1(c)). The PL spectrum of left (right) -circularly polarized component from the left-twisted gammadion sample matches to the PL spectrum of right (left) -circularly polarized component from the right-twisted sample. That is, the circularly anisotropy of the emitted light strongly depends on the handedness of a chiral photonic crystal. Degree of polarization of the emission,$(I_{LCP} - I_{RCP})/(I_{LCP} + I_{RCP})$, from left- and right-twisted samples are shown in Fig. 2(b). The maximum value of degree of polarization achieved is as large as 25.2\% at 1042nm for the right-twisted sample. 

\begin{figure}
 \includegraphics[width=80mm]{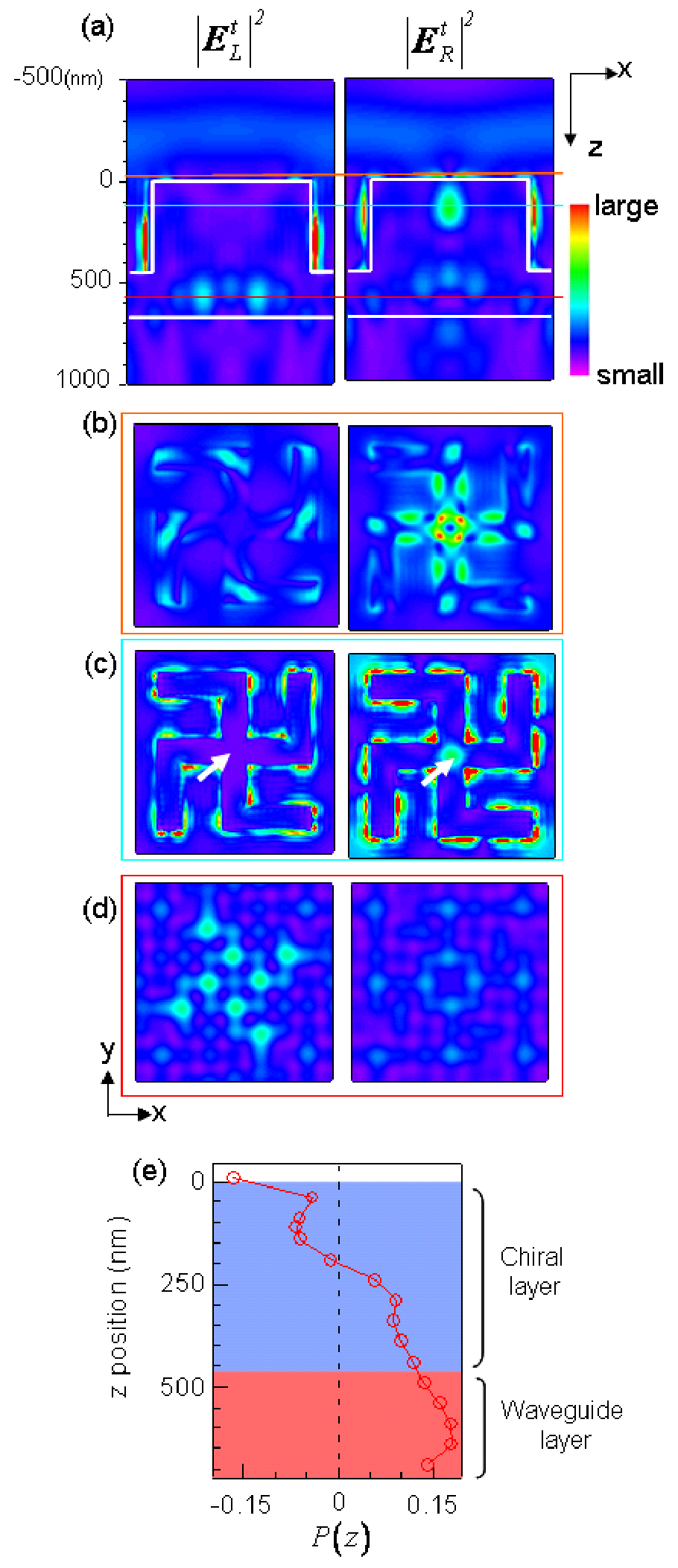}
 \caption{(a) Distribution of the electric field intensity induced by left- and right- circularly  polarized light ($|\Vec{E}_{tL}|^2$  and $|\Vec{E}_{tR}|^2$, respectively) at normal incidence tuned to 1040nm, sliced on the $xz$ plane including the center of the gammadion structure. The substrate is modeled as a semi-infinite slab of GaAs. The interfaces of GaAs layer are indicated by white line. Orange, blue and red lines indicate the position of the planes on which the calculations for (b)(c)(d) are preformed. (b)(c)(d) In-plane distribution of the electric field intensity induced by left- and right- polarized light at normal incidence. the $z$ position of the calculated planes for (b)(c)(d) are -10nm, 90nm, 590nm, respectively. (e) Dependence of the degree of polarization on the value of $z$. The definition of degree of polarization is explained in text. }
\end{figure}

The observed correlation between optical activity and polarization degree of the emission can be explained in terms of the strong left-right imbalance electromagnetic modes in chiral nanostructures. In linear optics, the electric field at a frequency $\omega$ at the point \Vec{r}, $\Vec{E}^t(\Vec{r})$, is proportional to the amplitude of the incident wave, $\Vec{E}^{in}(\Vec{r})$,
\begin{equation}
\Vec{E}^t_i(\Vec{r})=\int d\Vec{r}' L_{ij}(\Vec{r},\Vec{r}-\Vec{r}')\Vec{E}^{in}_j(\Vec{r}')
\end{equation}
where subscripts label Cartesian coordinates and $L_{ij}(\Vec{r},\Vec{r}-\Vec{r}')$ is a tensorial function that that depends on the material properties and geometry of the interaction. In an homogeneous medium, this function depends only on $\Vec{\rho}=\Vec{r}-\Vec{r}'$, however in the inhomogeneous medium, it is also a function of $\Vec{r}$. 
Spontaneous emission rate into a specific mode depends on the amplitude of the vacuum field of the relevant mode at the position of the emitter \cite{28}. If we consider a mode, which connects to plane wave propagating with a wave vector $\Vec{k}_0$ with a vacuum field amplitude of free space at large distance from the structure, the relevant Fourier component of the operator, 
\begin{equation}
l_{ij}(\Vec{r},\Vec{k}_0)=\int d\Vec{\rho} L_{ij}(\Vec{r},\Vec{\rho})e^{-i\Vec{k}_0\cdot \Vec{\rho}}
\end{equation}
can be a measure of the enhancement or suppression factor of spontaneous emission rate to that mode at the point of interest \Vec{r}. Thus one can restore the medium-induced modification of the vacuum field amplitude at \Vec{r} by measuring $l_{ij}(\Vec{r},\Vec{k}_0)$ for all $\Vec{k}_0$. When we are interested in the modification of the vacuum field, which are coupled with light propagating along the surface normal in planar structures, the restoration of the relevant mode structure is reduced to the light field distribution inside the structure produced by a plane wave, i.e. $\Vec{E}^{in}(\Vec{r})=\Vec{A}e^{ik_0 z}$  . If the planar structure is isotropic in the $XY$ plane, the symmetry reduces the operator $\hat{l}(\Vec{r},\Vec{k}_0)$ down to a constant, $\Vec{E}^t(\Vec{r})=l_0(\Vec{r})\Vec{A}$, that describes modification of the vacuum modes propagating along the slab normal. In an isotropic chiral layer, the counter circularly-polarized vacuum modes will be modified differently, i.e. $\Vec{E}^t_{L,R}(\Vec{r})=l_{L,R}(\Vec{r})\Vec{A}_{L,R}$  where subscripts $L$ and $R$ stand for left- and right-circular polarizations, respectively. $l_{L,R}(\Vec{r})$ will give us the spatial distribution of the enhancement or suppression factor of the amplitudes of the left- and right-circularly polarized vacuum field modes coupled to the plane wave modes with wave vector normal to the surface of the structure. Thus if a point dipole oscillator equally emits left- and right-circularly polarized photons in homogeneous media, the same dipole embedded into the chiral slab at point $\Vec{r}$ will emit radiation with a finite degree of polarization, in the case of $|\Vec{A}_L|=|\Vec{A}_R|$, given by 
\begin{eqnarray}
\xi (r)=\frac{|\Vec{E}^t_L(\Vec{r})|^2-|\Vec{E}^t_R(\Vec{r})|^2}{|\Vec{E}^t_L(\Vec{r})|^2+|\Vec{E}^t_R(\Vec{r})|^2}=\frac{|l_L(\Vec{r})|^2-|l_R(\Vec{r})|^2}{|l_L(\Vec{r})|^2+|l_R(\Vec{r})|^2}
\end{eqnarray}
Thus, circular anisotropy of the vacuum field $\xi(\Vec{r})$ in the chiral photonic structure can be visualized by comparing the light intensity at the point \Vec{r} when the structure is irradiated by left- and right-circularly polarized plane waves at normal incidence. When QDs are homogeneously distributed in the $XY$ plane at a given $z$ forming a two dimensional emission layer, the far-field plane wave emission normal to the surface shows the degree of polarization characterized by:
\begin{equation}
P(z)=\frac{\int \left[|l_L(\Vec{r})|^2-|l_R(\Vec{r})|^2\right]dxdy}{\int \left[|l_L(\Vec{r})|^2+|l_R(\Vec{r})|^2\right]dxdy}.
\end{equation}

Figure 3 shows calculation results of induced electric field distribution by left- and right- circularly polarized light with the same amplitude at normal incidence at the wavelength of 1040 nm featured by the strong ellipticity of the transmitted wave (see Fig. 1(c)). We used the rigorous coupled-wave analysis method \cite{29}. One can observe from Fig. 3(a) that there is a drastic difference in the light intensity distribution for counter-circularly-polarized incident waves, $|l_L(\Vec{r})|^2$ and $|l_R(\Vec{r})|^2$ . This difference visualizes the sensitivity of the vacuum field to the handedness of the circularly-polarized light in chiral photonic crystals.

Figure 3(b)(c) and (d) shows $l_{L,R}(\Vec{r})$ at $z$=-10nm (corresponds to the top of chiral layer), $z$=90nm (inside the chiral layer) and $z$=590nm (in the waveguide layer) and the obtained values of $P(z)$ are -16.5\%A-6.5\%A17.3\%, respectively. We can see that both in-plane spatial distribution of $l_{L,R}(\Vec{r})$ and degree of polarization in radiation zone $P(z)$  strongly depends on $z$, i.e.~on the position of an emission layer inside the chiral structure. One can observe from Fig. 3(e) that the largest degree of polarization can be achieved when the emission layer is placed at $z \approx$600 nm. The observed strong circular anisotropy in emission corresponds with the results obtained by the numerical calculation. The slightly bigger - than predicted by calculation - value of $P(z)$ obtained in the experiment is mainly due to the structure imperfectness.

In conclusion, we demonstrate a system that possesses left-right circularly-polarization asymmetry of the vacuum electromagnetic modes and observe a pronounced imbalance between left- and right-circular polarizations in light emitted by QDs embedded in the chiral photonic crystal. The obtained degree of polarization of PL spectra is as large as 25\%. Moreover, our analysis shows that the degree of polarization of light emitted by a QD can be controlled by choosing its position in the waveguide layer of the phonics crystal. For example, if we place a single emitter to the center of the gammadion which position shown by the white arrows in Fig. 2(c), the degree of polarization of the emitted light will be as large as 92\%. Such an optimal control of the circular polarization degree by introducing using left-right asymmetry in the vacuum field should be important for various applications including quantum information technology. These also lead to applications such as novel circularly polarized single photon and surface-emitting circularly-polarized laser sources. 

\begin{acknowledgments}
We are grateful to H.~Tamaru, Y.~Svirko and J.~B.~H\'{e}roux for fruitful discussion, and H.~Ono, T.~Sato and T.~Shimura for sample fabrication. This research was supported by the Photon Frontier Network Program, KAKENHI(20104002) and Special Coordination Funds for Promoting Science and Technology (SCF) of the Ministry of Education, Culture, Sports, Science and Technology, Japan.
\end{acknowledgments}


\begin{thebibliography}{99}
\bibitem{16} E. Yablonovitch, Phys. Rev. Lett. \textbf{58}, 2059 (1987).
\bibitem{17} Y. Yamamoto, and R. E. Slusher, Physics Today \textbf{46}, 66 (1993).
\bibitem{18} P. Lodahl \textit{et al.}, Nature \textbf{430}, 654 (2004).
\bibitem{19} D. Englund \textit{et al.}, Phys. Rev. Lett. \textbf{95}, 013904 (2005).
\bibitem{20} H. Altug, D. Englund, and J. Vu\u{c}kovi\`{c}, Nature Phys. \textbf{2}, 484 (2006).
\bibitem{21} E. Miyai \textit{et al.}, Nature \textbf{441}, 946 (2006).
\bibitem{22} S. Fan, P. R. Villeneuve, J. D. Joannopoulos, and E. F. Schubert, Phys. Rev. Lett. \textbf{78}, 3294  (1997).
\bibitem{23} S. Noda, M. Yokoyama, M. Imada, A. Chutinan, and M. Mochizuki, Science \textbf{293}, 1123 (2001).
\bibitem{24} S. Strauf \textit{et al.}, Nature photon. \textbf{1}, 704 (2007).
\bibitem{11} R. Fiederling \textit{et al.}, Nature \textbf{402}, 787 (1999).
\bibitem{12} Y. Ohno \textit{et al.}, Nature \textbf{402}, 790 (1999).
\bibitem{13} E. I. Rashba \textit{et al.}, Phys. Rev. \textbf{B 62}, R16267 (2000).
\bibitem{14} X. Jiang, \textit{et al.}, Phys. Rev. Lett. \textbf{94}, 056601 (2005).
\bibitem{15} M. Holub, and P. Bhattacharya, J. Phys. D: Appl. Phys. \textbf{40}, R179 (2007).
\bibitem{30} M. Nomura, \textit{et al.}, Opt. Express \textbf{14}, 6308 (2006).
\bibitem{25} B. Bai, Y. Svirko, J. Turunen, and T. Vallius, Phys. Rev. \textbf{A 76}, 023811(2007).
\bibitem{26} K. Konishi \textit{et al.}, Opt. Express \textbf{16}, 7189 (2008).
\bibitem{27} B. Bai \textit{et al.}, Opt. Express \textbf{17}, 688 (2009).
\bibitem{1} L. D. Barron,  \textit{Molecular Light Scattering and Optical Activity} (Cambridge Univ. Press, Cambridge, 1983).
\bibitem{2} H. Rau, Chem. Rev. \textbf{83}, 535 (1983).
\bibitem{3} Y. Inoue, Chem. Rev. \textbf{92}, 741 (1992).
\bibitem{4} M. Kroutvar \textit{et al.}, Nature \textbf{432}, 81 (2004).
\bibitem{5} J. Berezovsky \textit{et al.}, Science \textbf{314}, 1916 (2006).
\bibitem{6} H. Krenn, W. Zawadzki, and G. Bauer, Phys. Rev. Lett. \textbf{55}, 1510 (1985).
\bibitem{7} D. D. Awschalom, J. Warnock, and S. von Moln\={a}r, Phys. Rev. Lett. \textbf{58}, 812 (1987).
\bibitem{8} S. Noda, Science \textbf{314}, 260 (2006).
\bibitem{9} M. T. Hill \textit{et al.}, Nature Photon. \textbf{1}, 589 (2007).
\bibitem{10} M. A. Noginov \textit{et al.}, Nature \textbf{460}, 1110 (2009).
\bibitem{31} K. Sato, Jpn. J. Appl. Phys. \textbf{20}, 2403 (1981).
\bibitem{32} M. Kuwata-Gonokami \textit{et al.}, Phys. Rev. Lett. \textbf{95}, 227401 (2005).
\bibitem{28} R. Loudon, \textit{The Quantum Theory of Light} (Oxford Science Publication, Oxford, 1973)
\bibitem{29} M. G. Moharam, and T. K. Gaylord, J. Opt. Soc. Am. A \textbf{3}, 1780 (1986).

\end{thebibliography}
\end{document}